\documentclass{ws-ijmpcs}

\usepackage[active]{srcltx}
\usepackage{fontenc}
\usepackage{graphicx}
\usepackage{amsmath,amsfonts,bm,bbold}
\usepackage{amssymb}
\usepackage{slashed}
\usepackage{subfigure}

\def\trimmarks{}

\begin{document}

\markboth{A.\,Tandogan  and A.\,V.\,Radyushkin}
{Method of Analytic Evolution of Flat Distribution Amplitudes  in QCD}

%
\catchline{}{}{}{}{}
%

\title{METHOD OF ANALYTIC   EVOLUTION OF  FLAT DISTRIBUTION AMPLITUDES IN QCD\footnote{Preprint JLAB-THY-11-1433}}

\author{ASLI TANDOGAN\footnote{e-mail: atand002@odu.edu}  \    \ 
and ANATOLY V. RADYUSHKIN\footnote{Also at Bogoliubov Laboratory of Theoretical Physics, 
JINR, 141980 Dubna, Russian           Federation; e-mail: radyush@jlab.org}  }
\address{Physics Department, Old Dominion University, Norfolk,
             VA 23529, USA  \\ {\rm and} \\
Thomas Jefferson National Accelerator Facility,
             Newport News, VA 23606, USA }

\maketitle

\begin{history}
\end{history}

\begin{abstract}

A  new analytical method of performing   ERBL evolution is described.
The main goal is to develop an   approach    that works 
 for distribution amplitudes  that do not vanish at the end points, for which 
 the standard method of expansion  in Gegenbauer polynomials
 is inefficient. 
Two cases  of the  initial DA  are considered:   a purely flat DA,
given by the same constant    for all $x$, 
and an antisymmetric DA given  by opposite   constants  for $x \gtrless 1/2$. 
For  a purely flat DA,  the evolution is governed by an overall  $(x \bar x)^t$  
dependence on the evolution parameter $t$ times a factor   that was
 calculated as an expansion in $t$.
For  an antisymmetric flat DA, 
 an extra overall  factor $|1-2x|^{2t}$ appears due to a  
 jump at  $x=1/2$.  
A  good convergence was  observed in the $t \lesssim 1/2$
region. For larger $t$,  one can use the standard 
method of the Gegenbauer expansion.

\end{abstract}

\ccode{PACS numbers: 12.38.-t, 11.10.-z}

 \section{Introduction}

Recent BaBar data\cite{Aubert:2009mc}  on the $\gamma^* \gamma \to \pi^0$   
transition form factor   correspond to approximately logarithmic
$\ln Q^2$
raise of the combination $Q^2 F_{\gamma^* \gamma \pi^0} (Q^2)$
in the region of very  high momentum transfers 10 to 40 GeV$^2$,
where  perturbative QCD approach\cite{Lepage:1980fj}  predicts nearly constant behavior for this 
combination.   It was proposed\cite{Radyushkin:2009zg,Polyakov:2009je} to explain the BaBar ``puzzle'' 
by assuming that the pion distribution amplitude (DA) is ``flat'':
$\varphi_\pi (x) =f_\pi$. 

In general,  DAs $\varphi_\pi (x,\mu)$  depend on the normalization scale $\mu$,
tending to the ``asymptotic'' shape $\varphi_\pi^{\rm as}  (x) = 6 f_\pi x (1-x)$
in the $\mu \to \infty$ limit, for any intial pion DA.
A standard way\cite{Efremov:1979qk,Lepage:1980fj}  to this  result  
is to expand the initial pion DA  over  the  eigenfunctions $x(1-x)C_n^{3/2} (2x-1)$
of the evolution kernel.  Each    Gegenbauer projection  then changes 
as $[\ln \ln (\mu_0/\Lambda  )/\ln\ln (\mu/\Lambda)]^{\gamma_n /\beta_0}$
when  $\mu$ increases. All anomalous dimensions $\gamma_n$ are positive,
except for $\gamma_0$ which is zero,  hence only the $\sim x(1-x)$ part survives.
 For a pion DA  given by a sum of a few Gegenbauer polynomials,
this method gives a  convenient analytic expression for the DA evolution. 
However, to get a  flat pion DA, one should formally take an infinite number
of Gegenbauer polynomials (this  difficulty was mentioned 
	in a recent paper\cite{Brodsky:2011yv}), and one should take a very  large 
number of terms to get a reasonably precise  (point by point) 
result for the evolved DA.
In fact, a long  time ago,  it was  established\cite{Efremov:1979qk}  
that,  if the initial pion DA $\varphi_\pi (x,\mu_0)$ has 
a $\sim [x(1-x)]^r$ form near  the end points $x=0,1$, with a small $r$, then  evolution 
just increases this power: 
$$r\to r(\mu) = r(\mu_0) + 2 C_F \ln [\ln (\mu/\Lambda) / \ln (\mu_0/\Lambda)]/\beta_0 \  .$$
 This result was obtained by incorporating the large-$n$ behavior 
$\gamma_n \sim 4C_F \ln n$ of the anomalous dimensions in QCD, and 
summation of the leading end-point region terms.
Thus, one may expect that the evolved form 
of the flat pion DA should be close to 
$ [x(1-x)]^{2 C_F \ln [\ln (\mu/\Lambda) / \ln (\mu_0/\Lambda)]/\beta_0}$.
In particular,  this statement was repeated  in a 
more recent paper\cite{Broniowski:2007si}  on the basis of an analysis 
similar to that of Ref.[\refcite{Efremov:1979qk}]. 

Our goal  is to develop a method that  allows  one  to 
get explicit analytic  expressions  for the evolution of the flat distribution 
amplitude in QCD. As we  will see, our results  not only confirm the above expectation,
but also provide a  regular way of  calculating  corrections  to it.
We also apply this  new method to obtain the evolution of the 
DA  $d(x)  \sim {\rm sgn}   (x-1/2 )$   that is antisymmetric  with respect to
the central point $x=1/2$. Such DA may correspond  to the 
so-called ``D-term''\cite{Polyakov:1999gs}  in  generalized parton distributions.

\section{Evolution of Flat DA}

\subsection{Evolution equation}

Evolution of  the pion DA in the leading logarithm  approximation  is  governed  by  
\begin{eqnarray}
  \frac{\partial \varphi (x,t)}{\partial t}=\int_0^1  [V(x,y)]_+\,  \varphi(y,t)dy,
\end{eqnarray}
where $t=2C_F \ln\ln(\mu/\Lambda)/b_0$ is the 
leading logarithm 
QCD evolution parameter,   and $[V(x,y)]_+$ 
is the  evolution kernel which  has the property of   a  plus  distribution 
\begin{align}
 [V(x,y)]_+=V(x,y)-\delta(y-x)\int_0^1 V(z,y)dz 
\end{align}
with respect to its first argument. 
In our case, 
\begin{eqnarray}
V(x,y)=\left[\frac{x}{y}\left(1+\frac{1}{y-x}\right)\right]\theta(x<y)+
\left[\frac{\bar{x}}{\bar{y}}\left(1+\frac{1}{x-y}\right)\right]\theta(y<x)  \ ,
\end{eqnarray}
where  we use a standard notation  $\bar {x}=1-x$ and $\bar{y}=1-y$.
One may rewrite  the evolution equation as 
\begin{eqnarray}
  \frac{\partial \varphi (x,t)}{\partial t}=
 \int_0^1 \left [ V(x,y) \varphi(y,t)- V(y,x) \, \varphi(x,t) \right ]\, dy  \ .
\end{eqnarray} 
It is clear that the singularities  
of $V(x,y)$ and $V(y,x)$  at $x=y$ 
cancel each other  in the integrand above,   even though 
the subtraction there does not look like a ``+''-prescription with respect to 
the integration  variable. Adding and subtracting $ V(x,y)\varphi(x) $ in the integrand, 
we  obtain the equation 
\begin{eqnarray}
 \frac{\partial \varphi (x,t)}{\partial t}=
\int_0^1 V(x,y)[\varphi(y,t)-\varphi(x,t)]dy-\varphi(x,t)\int_0^1[V(y,x)-V(x,y)]dy 
\end{eqnarray}
in which the first  term  has the structure of the ``+''-prescription with respect to the
integration variable, so that $1/(x-y)$ singularity of $V(x,y)$ is cancelled 
by zero of $\varphi(y,t)-\varphi(x,t)$ at $x=y$.
The integral in the second term (call it $-v(x)$) is also finite.
Taking the Ansatz 
\begin{eqnarray}
 \varphi(x,t) = e^{t v(x)}\Phi(x,t)\ ,
\label{ansatz}
\end{eqnarray}
we obtain for $\Phi(x,t)$ the equation 
\begin{eqnarray}
 \frac{\partial{\Phi  (x,t) }}{\partial t} =\int_0^1 V(x,y)[e^{t [v(y)  -v(x)]}\Phi(y,t)-\Phi(x,t)]\, dy\, ,
\end{eqnarray}
which does not have the second term. 
The solution for  $ \Phi(x,t)$ may  be written  as a series in $t$ 
\begin{eqnarray}
 \Phi(x,t)=\sum_{n=0}^{\infty}\frac{t^n}{n!}\Phi_n(x)\,  , 
\label{expansion}
\end{eqnarray}
with the functions $\Phi_n(x) $  satisfying the recurrence  relation
\begin{eqnarray}
  \Phi_{n+1}(x)=\int_0^1  V(x,y) \left[\sum_{l=0}^n  \frac{n!}{(n-l)!\,l!}\Phi_l(y)
  \left[    v(y) - v(x) \right]^{n-l}-\Phi_n(x)\right] dy \  .
\end{eqnarray}

\subsection{Singular Part}

It is instructive to consider first an auxiliary situation when the evolution kernel
is  given by  the singular part  only 
\begin{eqnarray}
V^{\text{sing}}(x,y)&=&\left(\frac{x}{y(y-x)}\right)\theta(x<y)+(x\rightarrow \bar{x},\,\,\, y\rightarrow \bar{y}) 
\end{eqnarray}
of the QCD kernel.
In this case
\begin{eqnarray}
 v^{\text{sing}}( x) \equiv - \int_0^1[V^{\text{sing}}(y,x)-V^{\text{sing}}(x,y)]dy= 2+ \ln(x\bar{x}) 
\label{vsing}
\end{eqnarray}
and the recurrence  relation  is given by 
\begin{eqnarray}
  \Phi_{n+1}(x)=\int_0^1  V(x,y) \left[\sum_{l=0}^n
  \frac{n!}{(n-l)!\,l!}\Phi_l(y)\left(\ln\frac{y\bar{y}}{x\bar{x}}\right)^{n-l}
  -\Phi_n(x)\right]   dy \   ,
\end{eqnarray}
or,  explicitly for the first terms,  
\begin{eqnarray}
\Phi_1 (x) &=&  \int_0^1 V(x,y)[\Phi_0(y)-\Phi_0(x)]dy \, ,  \\
\Phi_2 (x) &=&   \int_0^1V(x,y) \left[\Phi_0(y)\ln\left( \frac{y\bar{y}}{x\bar{x}}\right)+\Phi_1(y )-\Phi_1(x)\right]dy  \,  , \\
 \Phi_3 (x)   &=&  \int_0^1V(x,y)\left  [\Phi_0(y)\ln^2\left(\frac{y\bar{y}}{x\bar{x}}\right)+2\Phi_1(y)\ln \left(\frac{y\bar{y}}{x\bar{x}}\right)
 +\Phi_2(y)-\Phi_2(x)   \right ]dy  \,   . \ \   \
\end{eqnarray}
In this  approximation,  we can write 
\begin{eqnarray}
\varphi^{\text{sing}}(x,t)&=&  (x\bar{x})^t \,  e^{2t} \,
\left (\Phi_0(x)+t \Phi_1(x)+\frac{t^2}{2!}\Phi_2(x)+\frac{t^3}{3!}\Phi_3(x)
+ \ldots 
\right ) \ .
\end{eqnarray}
If we take the flat DA for $t=0$, i.e., $\Phi(x,0)=1$, this gives 
\begin{eqnarray}
 \Phi_0 (x)&=&1   \,   \\
\Phi_1 (x)&=&0    \,    \\
\Phi_2 (x)&=& -2\ln x\ln\bar{x} \\
\Phi_3 (x)&=&3\ln(x\bar{x})\ln x \ln\bar{x}+2\left [\ln x \, \text{Li}_2(x) 
+\ln\bar{x}\, \text{Li}_2(\bar{x}) \right ] 
\nonumber\\ & & 
-4\, \left [ \text{Li}_3(x)   + \,  \text{Li}_3(\bar{x}) \right ]+8 \zeta(3)
\end{eqnarray}

The graphical results for the expansion components are given in Fig.1.
\begin{figure}[h]
\centering
\subfigure[]{
\includegraphics[height=4.5cm]{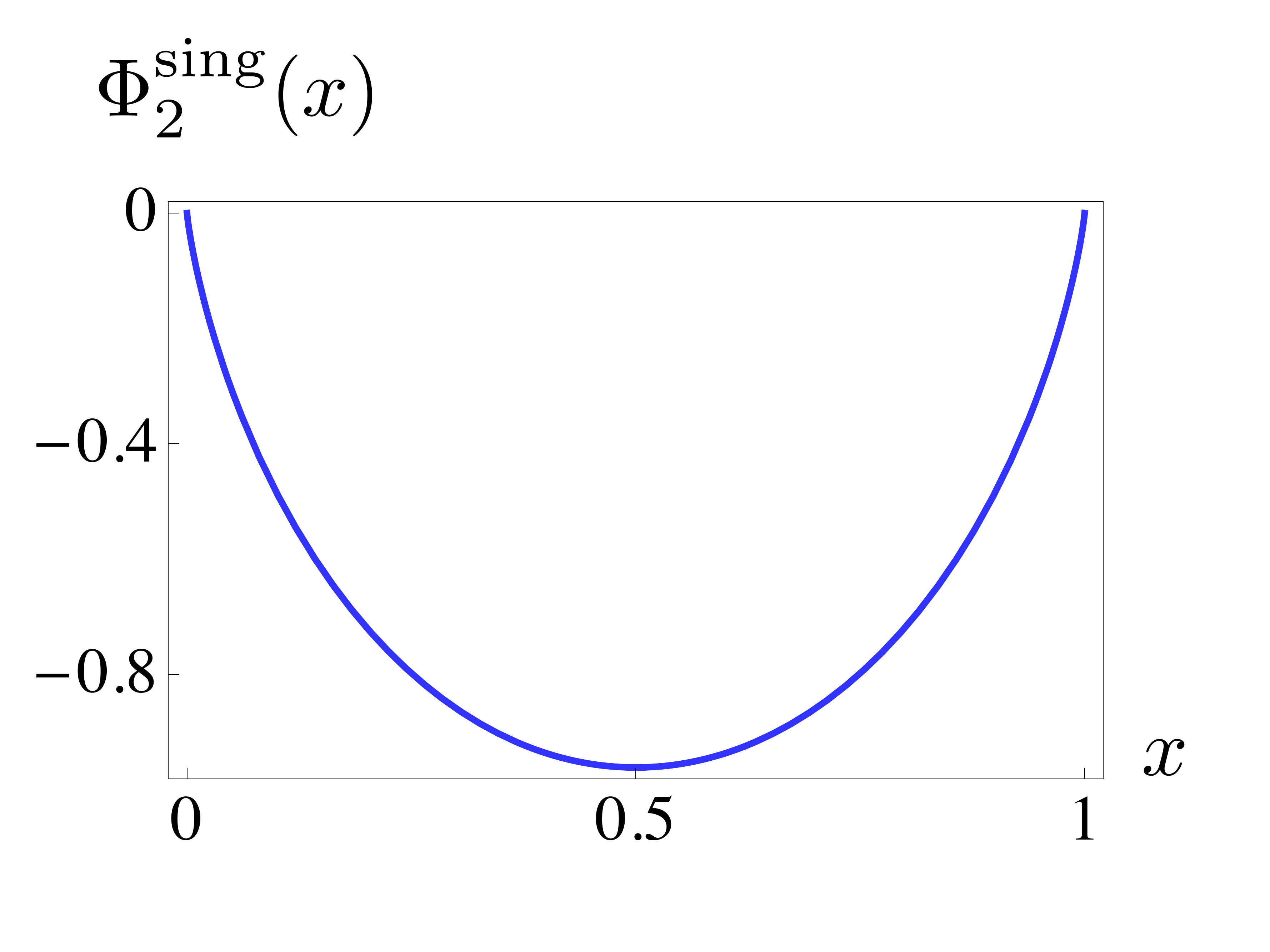}
\label{fig:subfig21}
}
\subfigure[]{
\includegraphics[height=4.3cm]{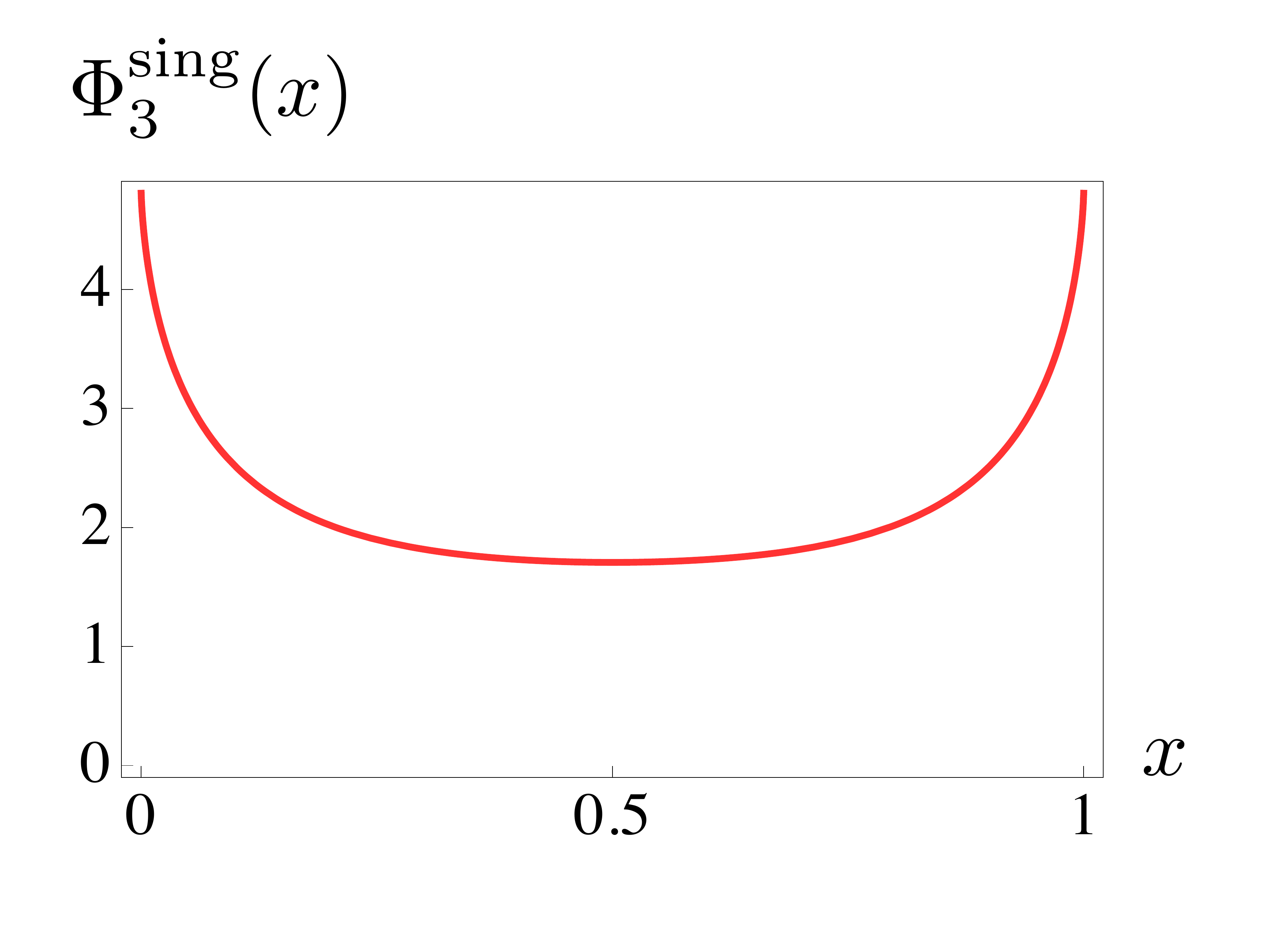}
\label{fig:subfig31}
}
\caption{Expansion components  $\Phi^{\text{sing}}_2(x)$ (a) and $\Phi^{\text{sing}}_3(x)$ (b). 
}
\end{figure}

As far as  $\int_0^1 [V(x,y)]_+ dx =0$, evolution does not change the normalization integral
for $\varphi (x,t)$. In particular, if we expand $\varphi (x,t)$  in  $t$
\begin{eqnarray}
\varphi (x,t)= \sum_{n=0}^\infty  \varphi_n (x)\,  \frac{t^n }{n!} \ ,
\end{eqnarray}
we should have
\begin{eqnarray}
 \int_0^1  \varphi_n (x) \, dx = \delta_{n0}  \ .
\end{eqnarray}
However, in our Ansatz $ \varphi(x,t)=e^{t(2+\ln(x\bar{x}))}\Phi(x,t)$, the $(t\ln x\bar{x})^N$ terms are summed to all orders, while the series over $\varphi_n(x)$ is restricted to some   finite order  $N$. 
As a result, the approximants $\varphi_{(N)}(x,t)$  are 
not normalized to 1.
In particular, if we keep the terms up to $\Phi_2 (x)$, the normalization integral  is  given  by
\begin{eqnarray}
I_2 (t) \equiv \int_0^1 \varphi_{(2)}^{\text{sing}}(x,t) \, dx=e^{2 t}  \left. \left. \frac { \Gamma^2(1+t))}{  \Gamma(2+2t) } 
\right  \{ 1-t^2 \left[(H_t-H_{1+2 t})^2-\psi_1(2+2 t)\right] 
  \right\} \  \  \ 
\end{eqnarray}
with  $H_n$ being  harmonic numbers and $\psi_k$  the  polygamma function.
One can check that
$I_2 (t) =1 + {\cal O} (t^3)$. 
For the next approximation, i.e., for $\varphi_{(3)}^{\text{sing}}(x,t) $,
the normalization integral is \mbox{$I_3 (t) =1 + {\cal O} (t^4)$,} etc.
For   $\varphi_{(N\rightarrow \infty)}(x,t)$, the normalization integral $I_N(t) $ will tend to 1 for all $t$. 

The normalization integrals versus $t$ are shown in Fig.2. 
For approximations involving $\Phi_0 (x)$ and  $\Phi_2 (x)$,  the calculations
were done  analytically,  while the curve  corresponding to inclusion of  $\Phi_{3}^{\text{sing}}(x,t) $
was  calculated numerically. 
As   seen from   Fig.2, adding more terms brings normalization closer to $1$.
\begin{figure}[h]
\centering
\includegraphics[width=8cm]{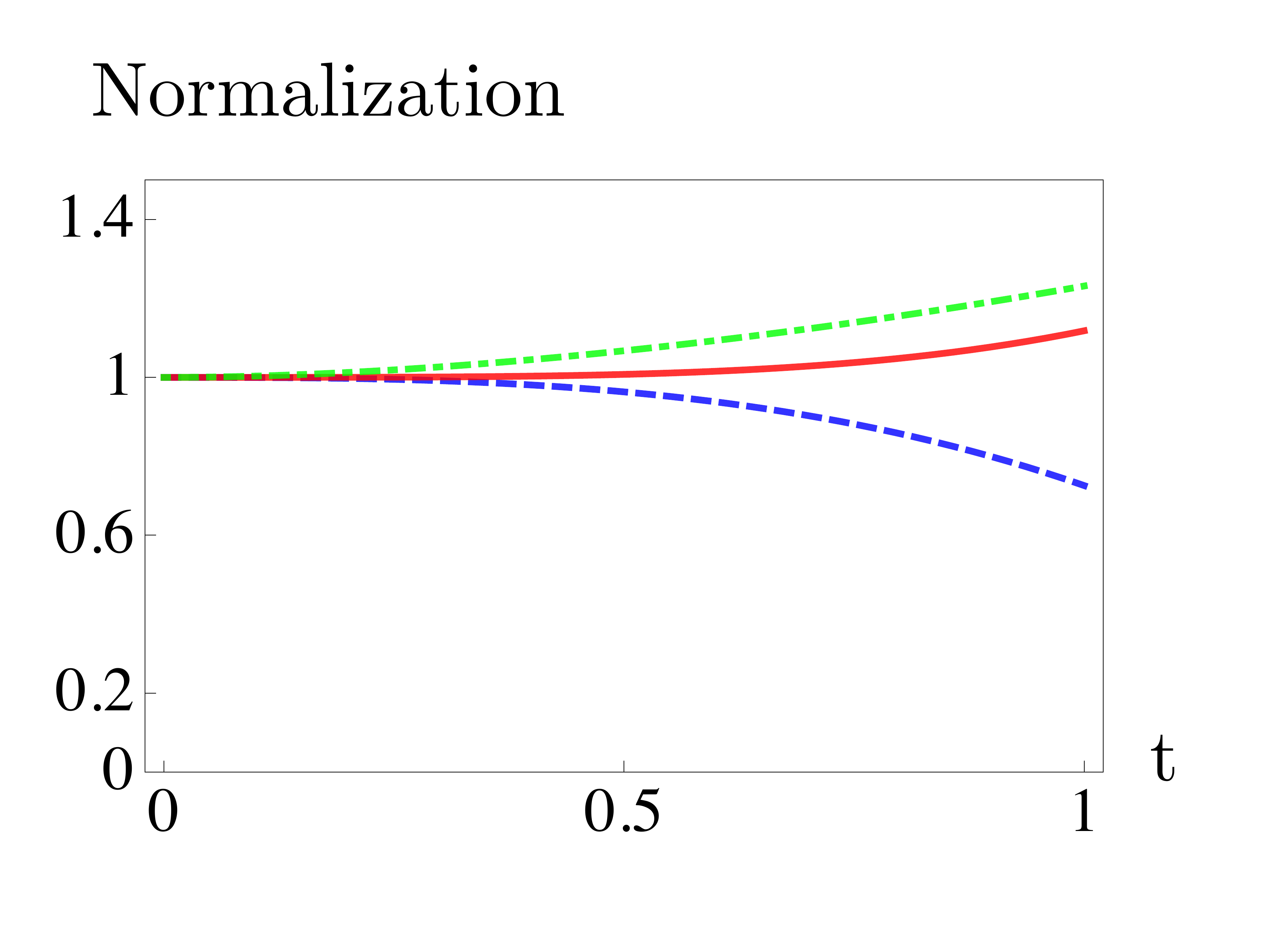}
\caption{ 
Normalization factor calculated for terms  including only $\Phi_0(x)$ (dot-dashed line), 
 $\Phi_0(x)$ and $\Phi_2(x)$ (dashed line) and  $\Phi_0(x)$, $\Phi_2(x)$ and $\Phi_3(x)$ 
   (solid line).}
\end{figure}

In this  situation, it makes sense to introduce the ``normalized Ansatz'', in which $\varphi (x,t)$ 
is approximated by the ratio $\nu_N (x, t) \equiv \varphi_{(N)}(x,t)/I_N(t)$, so that 
the correct normalization of the $N$th approximant is guaranteed for all $t$.
In particular, this gives
\begin{align}
 \nu_2 (x, t) =  (x\bar{x})^t \, \frac{\Gamma(2+2t) }{\Gamma^2(1+t)} 
\frac{1 -t^2 \ln x \, \ln \bar x   }{ 1-t^2 \left[(H_t-H_{1+2 t})^2-\psi_1(2+2 t)\right] }  \ .
\end{align}
As seen from this formula (and also from Fig.3), the initial flat 
 function immediately (for whatever small positive $t$)
evolves into a function vanishing 
 at the end points with its  shape dominated by  the $(x\bar{x})^t$
factor.

 \begin{figure}[h!]
\centering
\includegraphics[width=9cm]{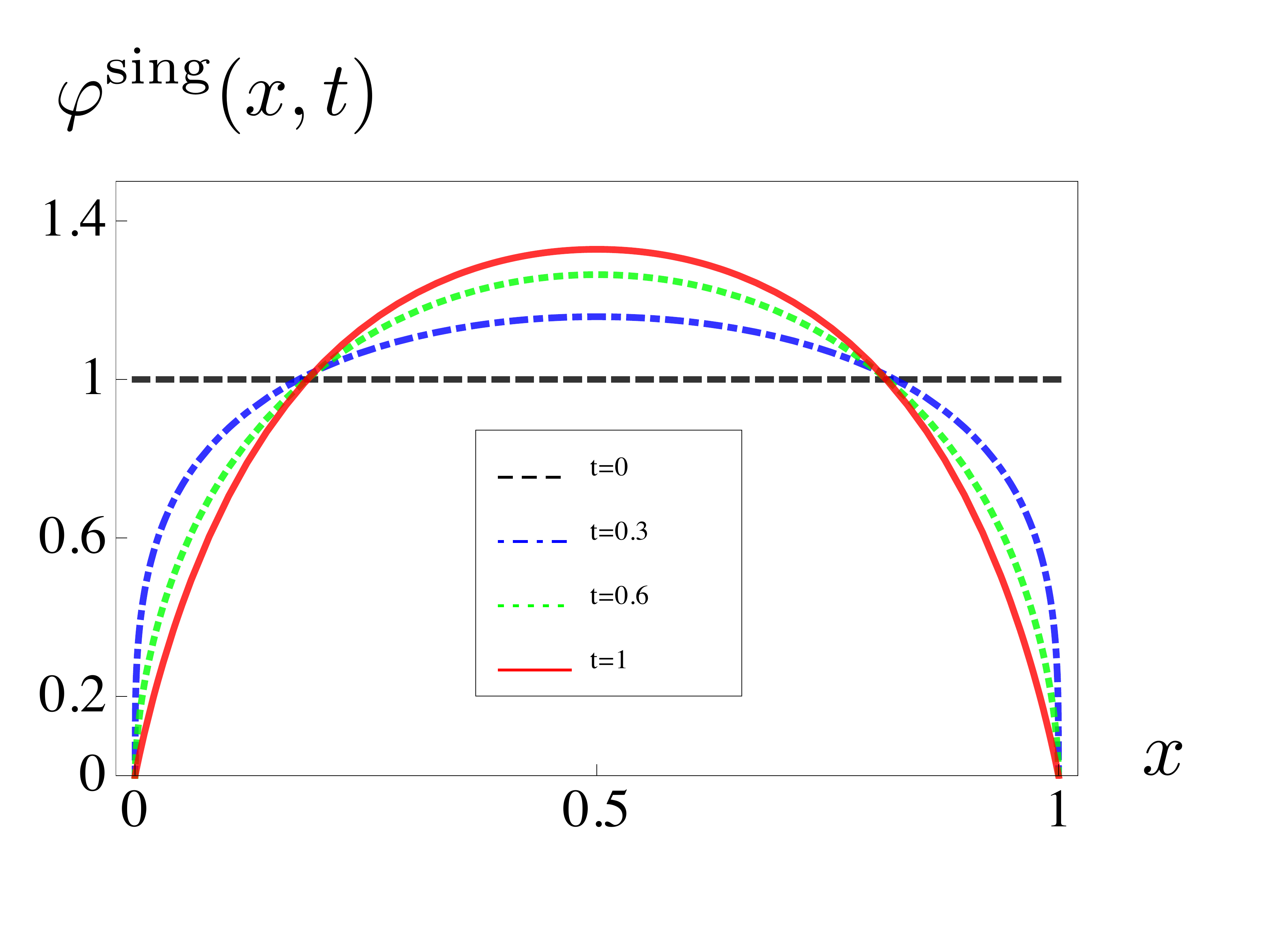}
\caption{Evolution of  the flat DA  under the singular part of the evolution kernel: 
the curves shown  correspond to $t=0, t=0.3, t=0.6, t=1.0$.}
\end{figure}

\subsection{Adding  Non-Singular Part}

\vspace{0.5cm}

When the whole QCD evolution kernel is taken into account, we have 
\begin{eqnarray}
 v(x) \equiv - \int_0^1[V(y,x)-V(x,y)]dy&=&3/2+x\ln\bar{x}+\bar{x}\ln x\\
&=&v^{\text{sing}}(x)-1/2 -x\ln x-\bar{x}\ln\bar{x}  \  .\nonumber  \label{nonsingular} 
\end{eqnarray}

Following the same steps as  in the previous section,  we  calculate the expansion 
terms  for the initial flat distribution amplitude: 
\begin{align}
 \Phi_0(x)=&1\\
 \Phi_1(x)=&0 \\
 \Phi_2(x)
=&
{x}\ln x+  \bar x \ln\bar{x} 
+\frac{1}{2} x \ln^2 x +\frac{1}{2} \bar{x}\ln^2 \bar{x}  +x\bar{x} \ln^2\frac{\bar{x}}{x}
-\ln x \ln\bar{x}
\nonumber\\ & + x\text{Li}_2\left(- \frac{\bar x}{x}\right)+
\bar{x}\text{Li}_2\left(-\frac{x}{\bar x}\right) 
 \ .
\end{align}
With these terms  taken into account  we have
\begin{eqnarray}
\varphi_{(2)}  (x,t)&=&  e^{3t/2} \, (x^{\bar{x}}\bar x^x)^t  
\left (1+\frac{t^2}{2!}\Phi_2(x)
\right ) \ .
\end{eqnarray}
Unfortunately,  for this form it is  impossible to analytically 
calculate the normalization integral even for the lowest term.
Compared  to the Ansatz used for the singular part of the
evolution kernel,    the  Ansatz
\begin{eqnarray}
 \varphi(x,t)=e^{t v(x) }  \Phi(x,t)=e^{3t /2 } 
(x^{\bar{x}}\bar x^x)^t \Phi(x,t)
\end{eqnarray}
has an extra overall factor 
$
[e^{-1 /2 }x^{-x} \bar x^{-\bar x}]^{t}  \ .
$
Note that the function
$x^{-x} \bar x^{-\bar x} $ 
is finite at the end points $x=0, \, 1$, where it 
takes its minimal value for the interval $[0,1]$
(equal to  1), 
and has a maximum for $x=1/2$, where it  equals 
2. Thus, the factor $
e^{-1 /2 }x^{-x} \bar x^{-\bar x}
$
enhances the $x\bar x$ profile in the middle 
(by $2/\sqrt{e}\approx 1.2$ factor)
and suppresses it at the end points (by $1/\sqrt{e}\approx 0.6$).
This is a rather mild  modification, and what is most important,
it does not change the $\sim x^t$ (or $\sim \bar x^t$) behavior at the end points.
So,  it makes sense to  use the expansion 
\begin{align}
 [x^{-x} \bar x^{-\bar x}]^{t}= \sum_{n=0}^\infty (-1)^n 
(x \ln  x + \bar x \ln \bar x)^n \, \frac{t^n}{n!} \ ,
\end{align}
in powers of $t$ 
and combine it with the expansion  for $\Phi (x,t)$.
This corresponds to Ansatz
\begin{eqnarray}
\varphi (x,t)&=&  (x\bar{x})^t \,  e^{3t/2} \,
\left (\tilde \Phi_0(x)+t \,  \tilde \Phi_1(x)+\frac{t^2}{2!}\tilde \Phi_2(x)+ \ldots 
\right ) \ ,
\end{eqnarray}
whose  expansion coefficients $\tilde \Phi_n (x)$ can  be straightforwardly 
obtained  from $ \Phi_n (x)$'s. 
In particular, $\tilde \Phi_1 (x) = -(x \ln  x + \bar x \ln \bar x)$,
and $\tilde \Phi_2 (x) = \Phi_2 (x) +  (x \ln  x + \bar x \ln \bar x)^2$.

\begin{figure}[t]
\centering
\subfigure[]{
\includegraphics[height=4.5cm]{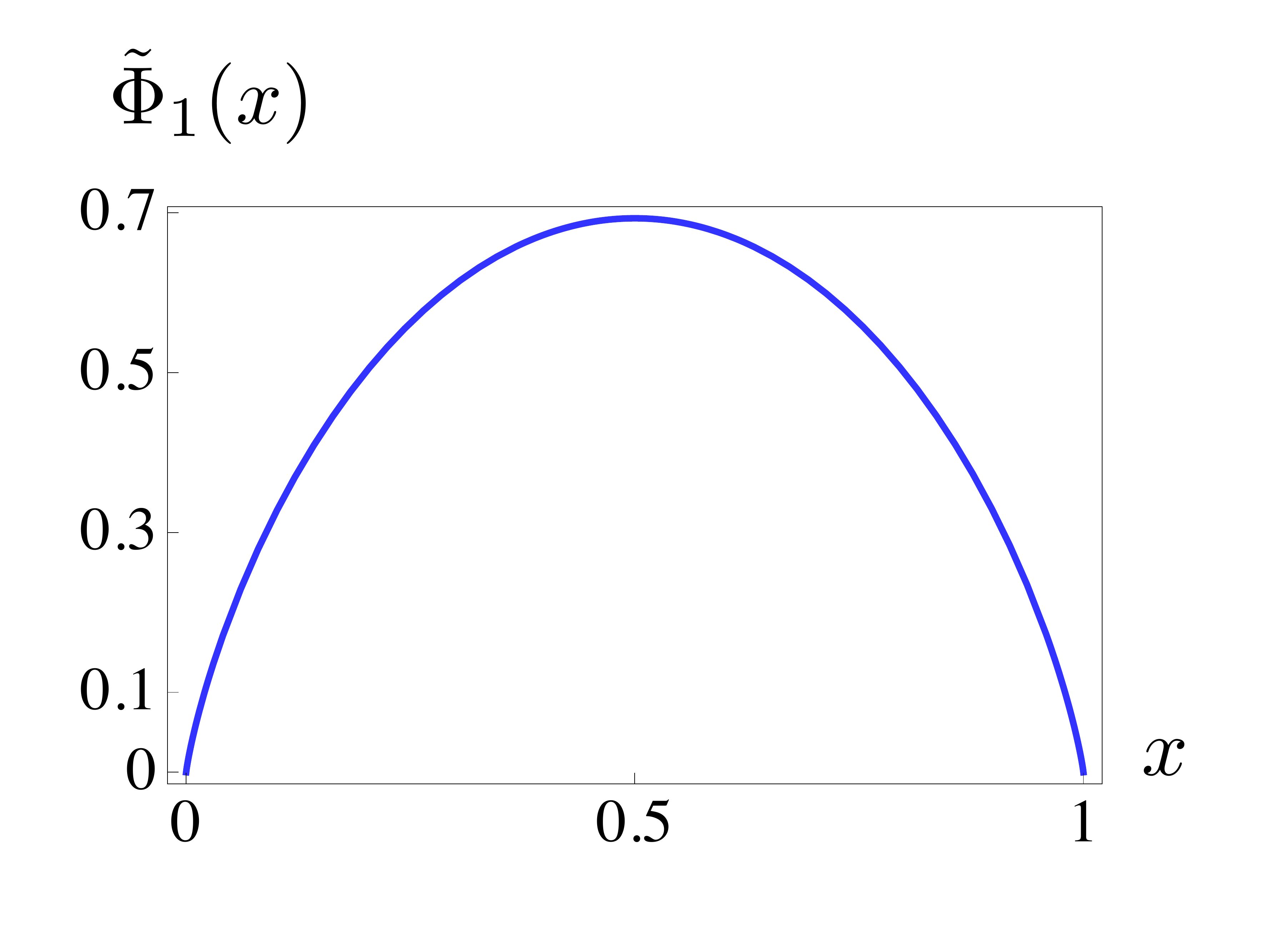}
\label{fig:subfig12}
}
\subfigure[]{
\includegraphics[height=4.3cm]{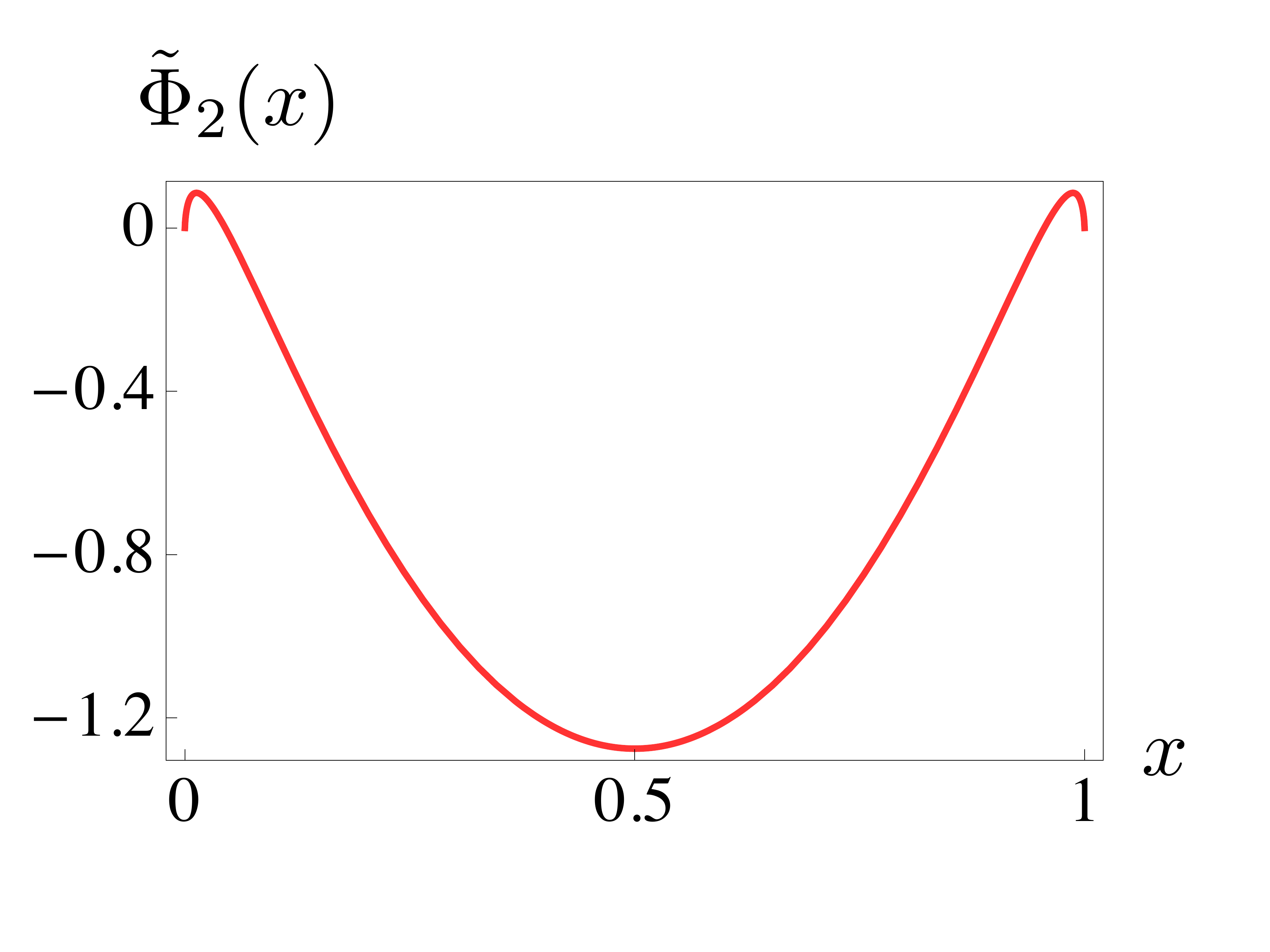}
\label{fig:subfig22}
}
\caption{Expansion components in the  full kernel case: $\tilde \Phi_1(x)$ (a) and $\tilde \Phi_2(x)$ (b). }
\end{figure}

 Now, the normalization integral for the lowest terms can be calculated analytically:
\begin{eqnarray}
I_1(t) \equiv \int_0^1 \varphi_{(1)} (x,t) dx=e^{3 t/2} 
\frac{ \Gamma^2(1+t)}{\Gamma(2+2t)} \left  (1 -\frac{t}{2(t+1)} + \frac{t}{2}  (-H_t+H_{1+2 t})\right ) \!
  ,
\end{eqnarray}
where  $H_n$ are harmonic numbers. 
Fig.5 shows the normalization versus $t$.

\begin{figure}[ht]
\begin{minipage}[b]{0.43\linewidth}
\centering
\includegraphics[width=6.5cm]{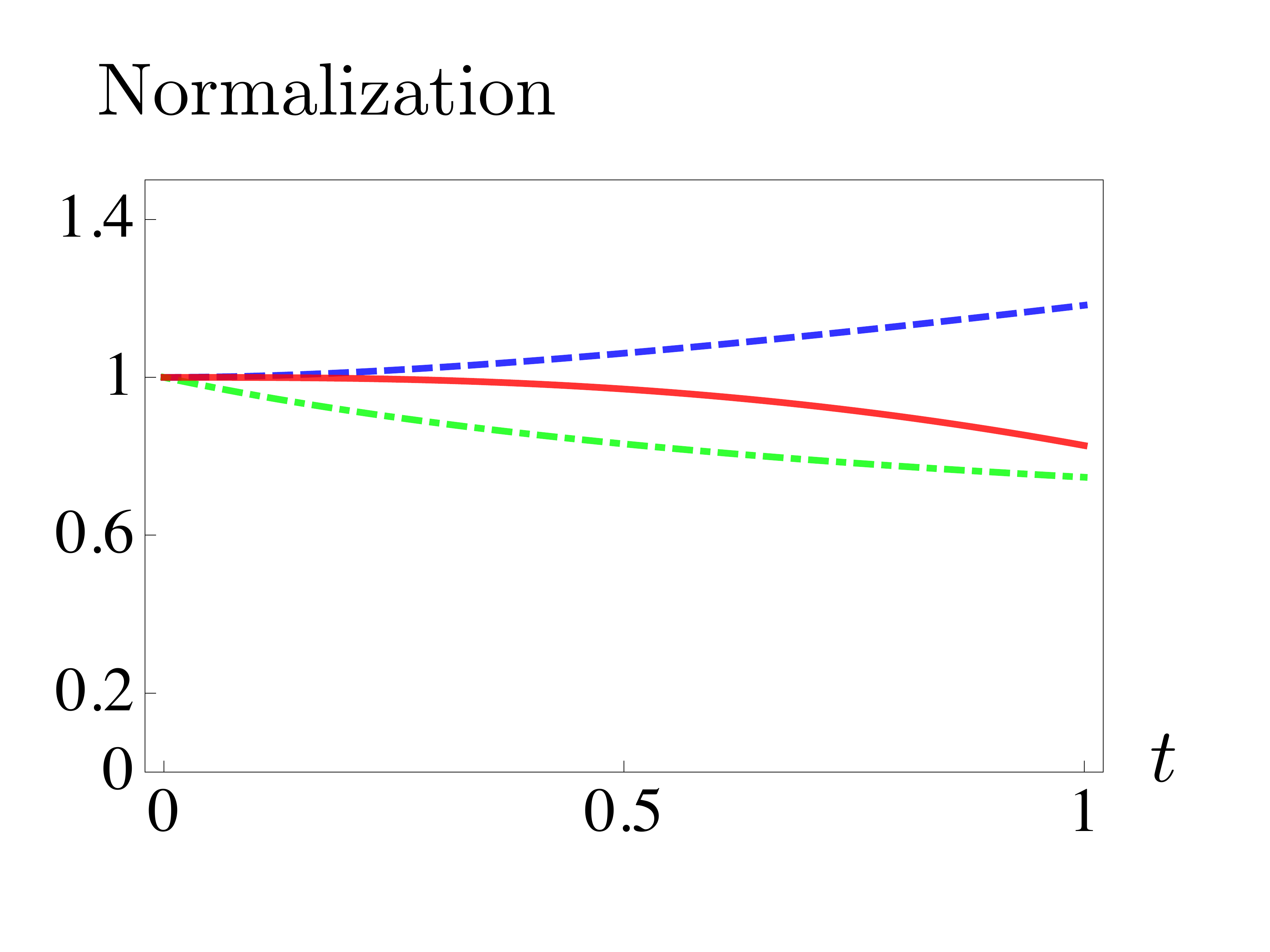}
\caption{Same as in Fig.2 for the case of the  full kernel.}
\end{minipage}
\hspace{1.2cm}
\begin{minipage}[b]{0.43\linewidth}
\centering
\includegraphics[width=6.5cm]{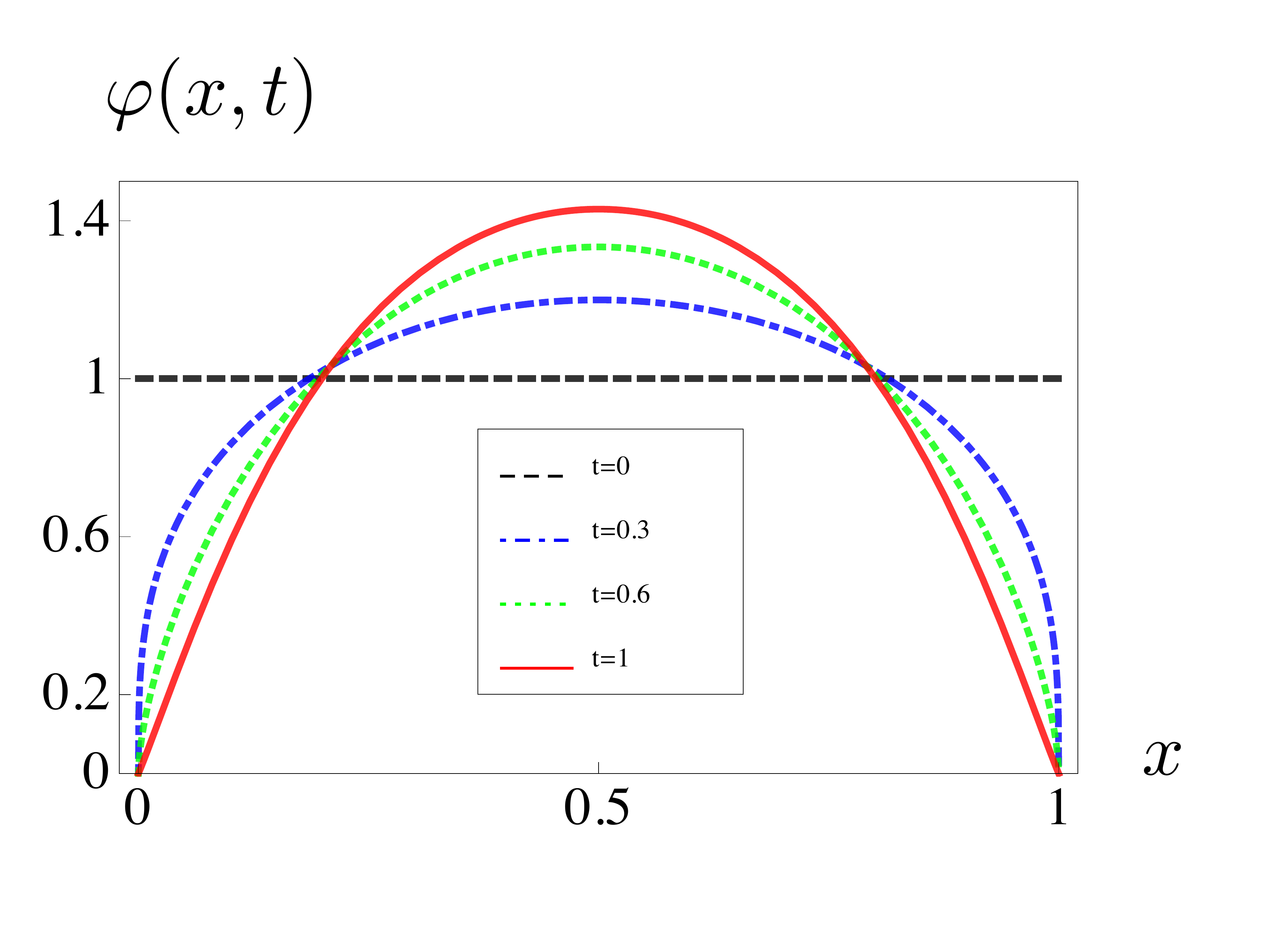}
\caption{Same as in Fig.3 for the case of the  full kernel.}
\end{minipage}
\end{figure}

Again, we may switch to the normalized Ansatz   formed by the ratio $ \varphi_{(N)} (x,t)/I_N(t)$.
For a flat initial distribution, this gives
\begin{align}
 \nu_1 (x, t) =  (x\bar{x})^t  \, \frac{\Gamma(2+2t) }{\Gamma^2(1+t)} 
\frac{1 -t \, (x \ln  x + \bar x \ln \bar x)  }{ 1-{t}/{2(t+1)} + {t}(-H_t+H_{1+2 t})/2  }  \ .
\end{align}
The results are illustrated by   Fig. 6.

\newpage

\section{Evolution of Anti-Symmetric Flat DA}

\vspace{0.3cm}

\subsection{Singular Part}

\vspace{0.3cm}

Evolution equations  may be applied also in situations when the distribution amplitude
is  antisymmetric with respect to the change $x \to 1-x$.
An  interesting  example is the $D$-term  function $d(x)$  that appears in 
generalized  parton  distributions. Thus,  let us  consider evolution of the DA  that
initially has the form 
\begin{displaymath}
\varphi_0(x)=\begin{cases} \,\,\,\,1 & 0<x\leqslant 1/2   \ , \\
 -1 & 1/2<x<1  \ .
\end{cases}
\end{displaymath}
The $v(x)$  function (\ref{vsing})
 is the same, since it depends on the kernel only.
Thus we can use the same Ansatz (\ref{ansatz})  and expansion (\ref{expansion}).
Since $\varphi_0(x)$ is  not just a constant, the first expansion coefficient $\Phi_1 (x)$
is  nonzero. 
Let us  start with the singular part of the kernel.
Then we get
\begin{displaymath}
 \Phi_1(x)=\begin{cases} -2 \ln\left[\frac{\bar{x}}{1-2 x}\right] & 0<x\leqslant 1/2  \ , \\
\,\,\,\,\,2 \ln\left[\frac{x}{-1+2 x}\right] & 1/2<x\leqslant 1  \ .
              \end{cases}
\end{displaymath}
We see that there  are logarithmic terms $ \ln |1-2 x|  $
singular for $x=1/2$. These    terms  are natural, since  each   half 
of the antisymmetric DA on its interval is expected to
 evolve similarly to a  flat DA
on the $0 \leq x\leq1$ interval. 
This observation suggests the Ansatz 
\begin{eqnarray}
 \varphi(x,t)=  e^{2t} (x\bar{x})^t |1-2x|^{2t}\Phi(x,t) \ .
\end{eqnarray}
With this definition of $\Phi (x,t)$, the   \mbox{$\ln|1-2x|$}  terms are eliminated from
 $\Phi_1(x)$: 
\begin{eqnarray}
 \Phi_1(x)=-2\ln\bar{x}  \ \theta ( 0<x\leqslant 1/2) - \{x \to \bar x \} \ .
    \end{eqnarray}    
          
For  the expansion component $\Phi_2(x)$, we have
\begin{align}
 \Phi_2(x)=& \theta ( 0<x\leqslant 1/2)
 \Biggl \{
-\frac{2\pi ^2}{3}-2\ln\bar{x}\ln x-\ln^2(1-2x)-3\ln\bar{x}(\ln(x\bar{x})-1)\nonumber\\
& -2\ln(1-2x)(\ln(4\bar{x}x)+
\ln(x\bar{x}))-8\text{Li}_2(1-2x) -4\text{Li}_2(x)-4\text{Li}_2(2x) 
\nonumber\\   &+2\text{Li}_2\left[\frac{x}{\bar{x}}\right]
+4\text{Li}_2\left[\frac{x}{2x-1}\right]+4\text{Li}_2\left[\frac{1-2x}{\bar{x}}\right]
\Biggr \} - \{x \to \bar x \} \   .
\end{align}

 The graphical results for the expansion components are shown in Fig. 7.
 
\begin{figure}[h]
\centering
\subfigure[]{
\includegraphics[height=4.5cm]{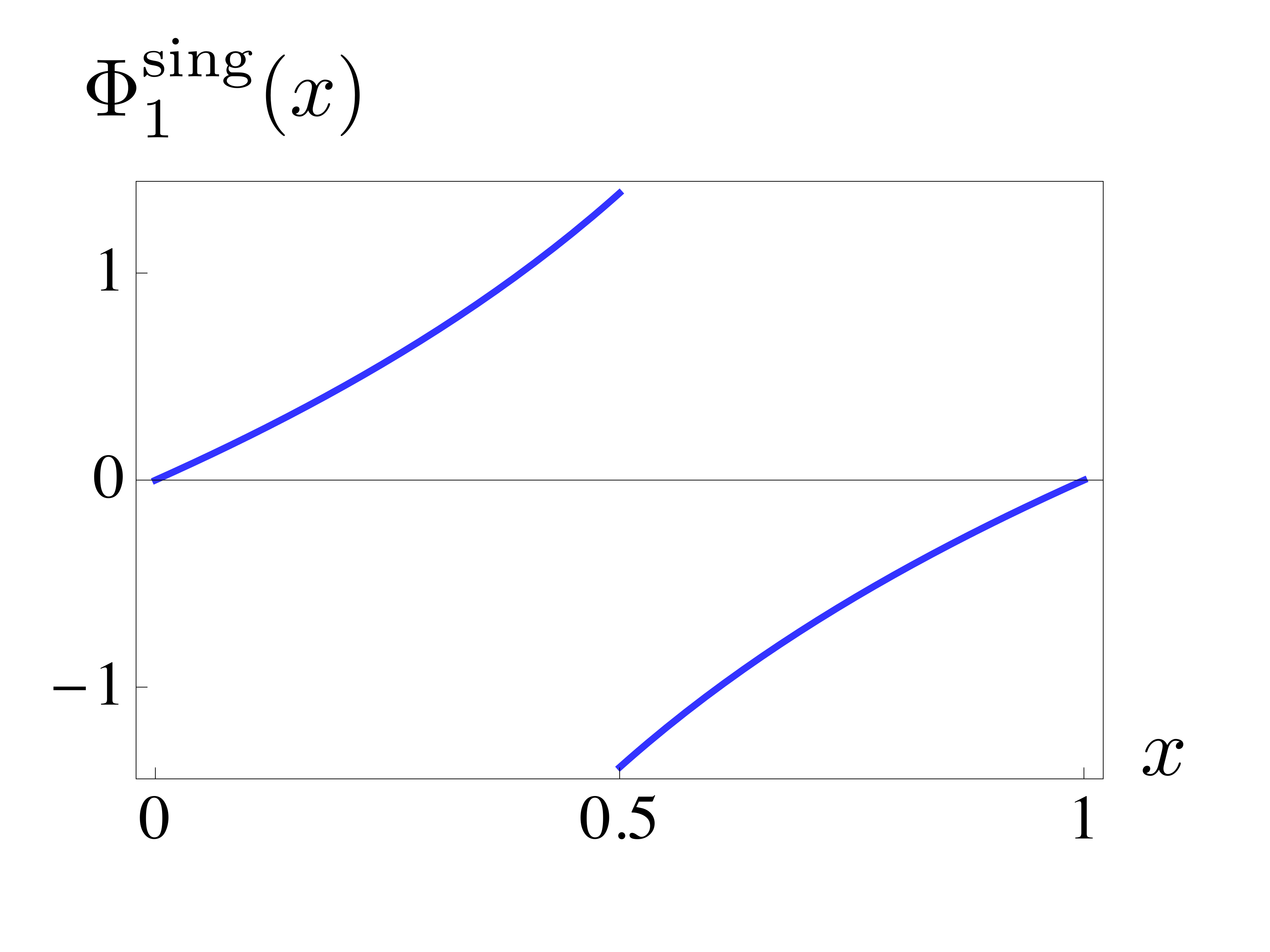}
\label{fig:subfig24}
}
\subfigure[]{
\includegraphics[height=4.5cm]{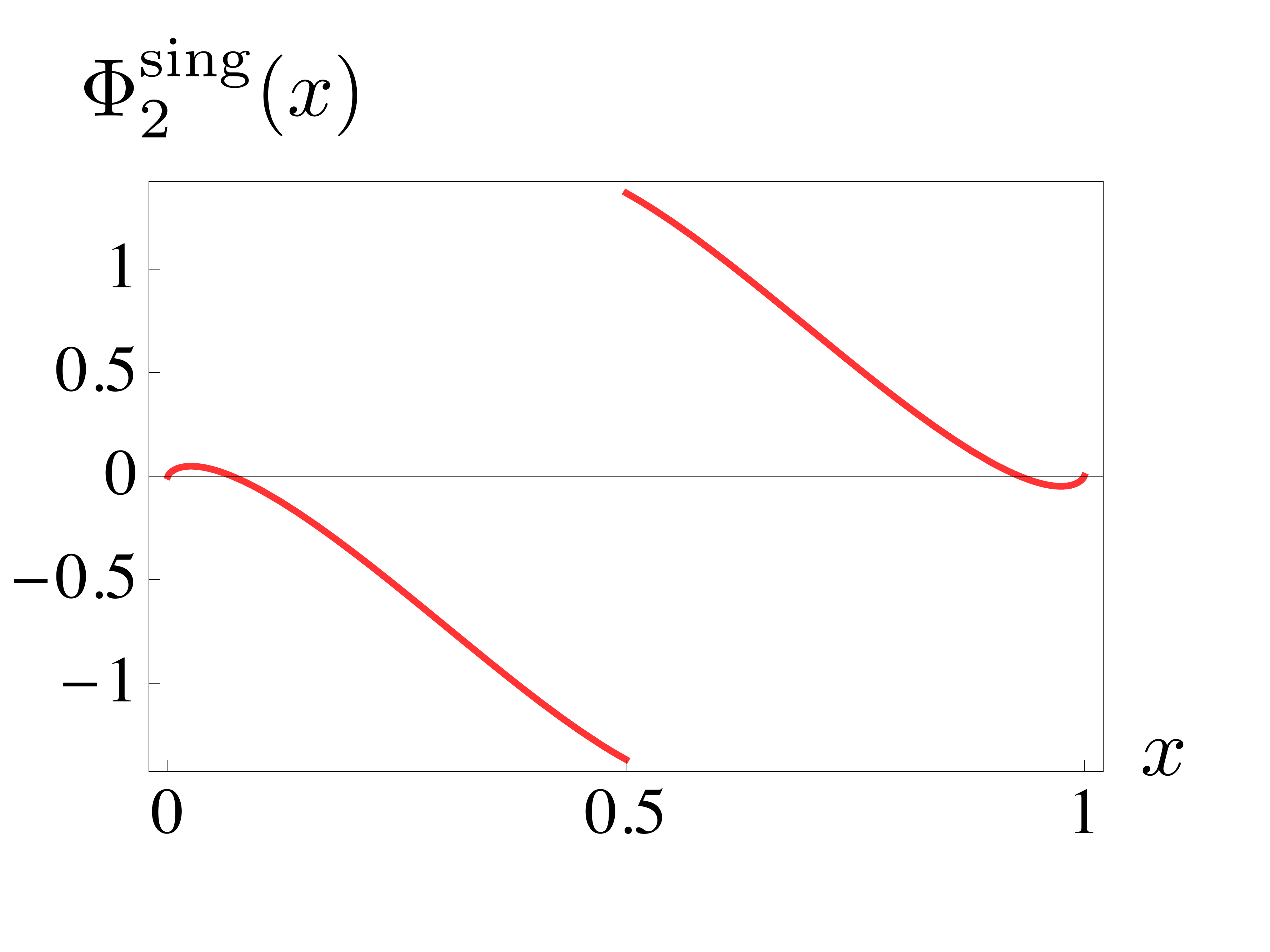}
\label{fig:subfig34}
}
\label{fig:subfigureExample}
\caption{Expansion coefficients for the antisymmetric DA:  $\Phi^{\text{sing}}_1(x)$ (a), 
$\Phi^{\text{sing}}_2(x)$ (b).  }
\end{figure}

The evolution of $\varphi_{\text{sing}}(x,t)$ to this accuracy can be obtained from 
\begin{eqnarray}
 \varphi(x,t)= e^{2t} (x\bar{x})^t|1-2x|^{2t}\left(\varphi_0(x)+t \Phi_1(x)+\frac{t^2}{2!}\Phi_2(x)\right) .
\end{eqnarray}

As  shown in Fig.8, the initial step function evolves into a function which is zero at the end  points
and in   the middle point.
\begin{figure}[h!]
\centering
\includegraphics[width=9cm]{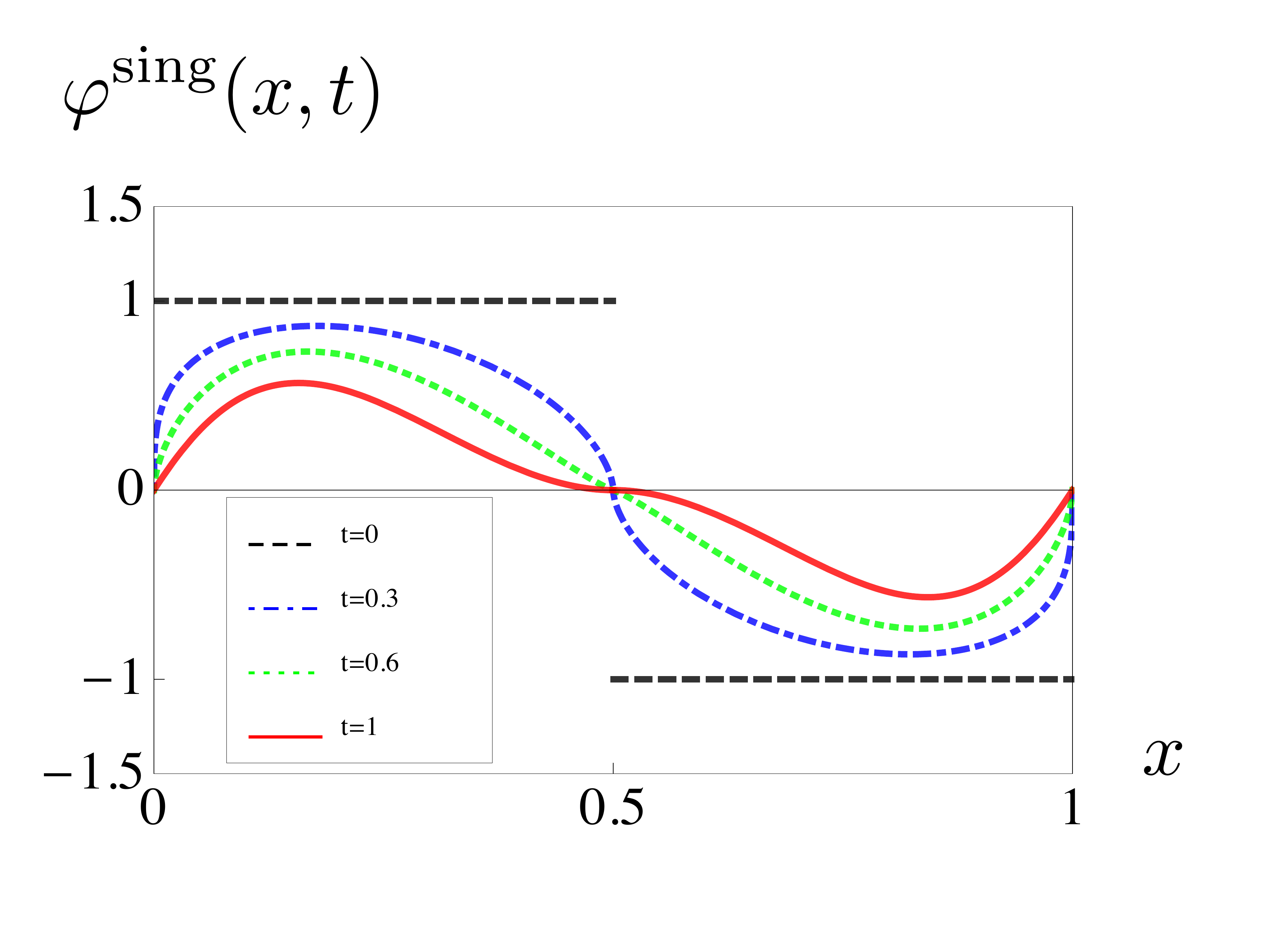}
\caption{Evolution of antisymmetric DA under the singular part of the kernel for  
$t=0, t=0.3, $ $ t=0.6, t=1$
}
\end{figure}

\newpage 

\subsection{Adding the Non-Singular Part}

\vspace{0.3cm}

Since the  nonsingular part  does  not add $\ln x$  and $\ln \bar x$  terms to  $v(x)$, 
we may  proceed with the same Ansatz 
\begin{eqnarray}
 \varphi(x,t)=&e^{3t/2} (x\bar{x})^t |1-2x|^{2t} \Phi(x,t)\, ,
\end{eqnarray}
but the expansion components change (see Fig.9).
\begin{figure}[hb]
\centering
\subfigure[]{
\includegraphics[width=6cm]{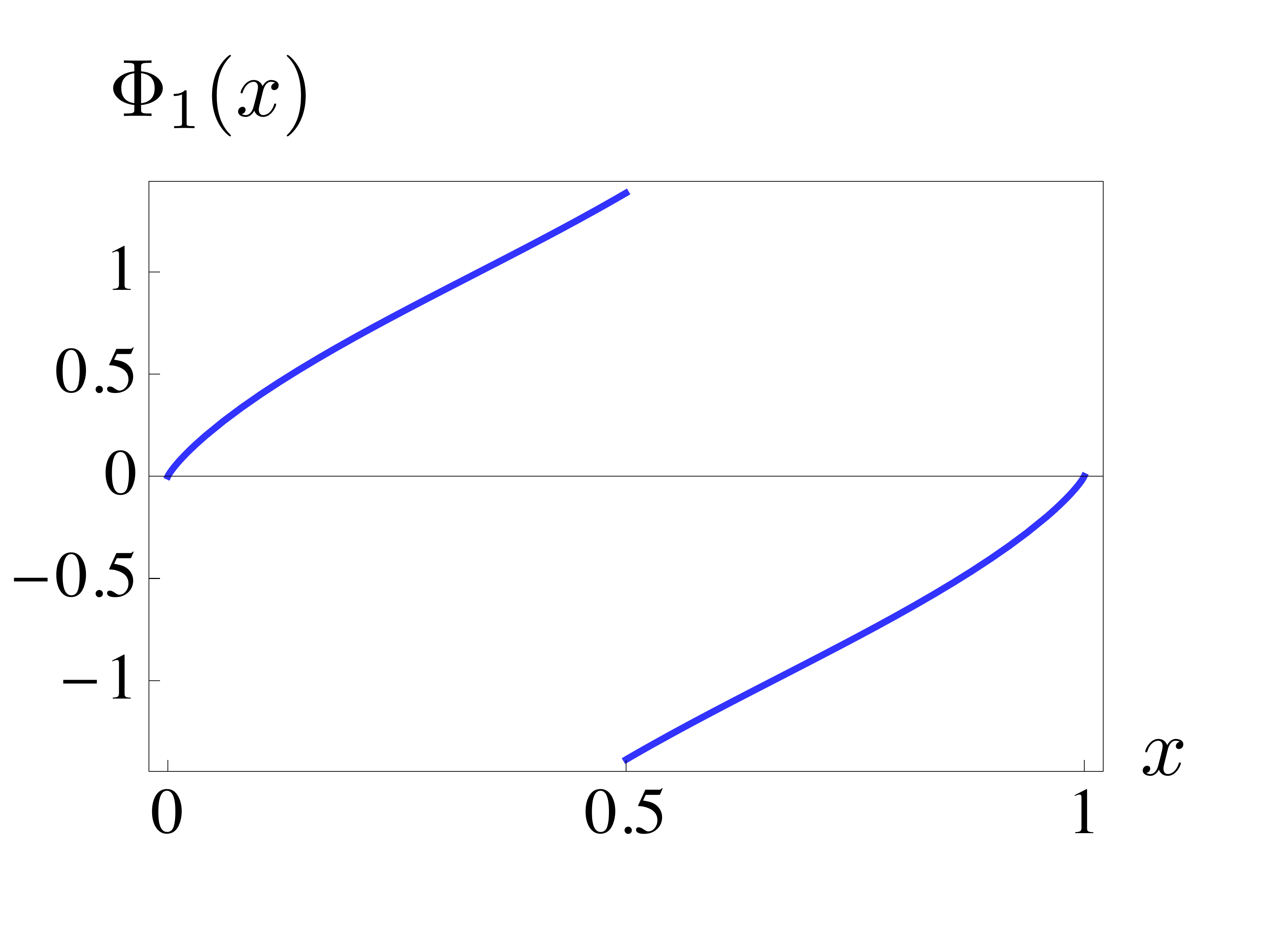}
\label{fig:subfig25}
}
\subfigure[]{
\includegraphics[width=6cm]{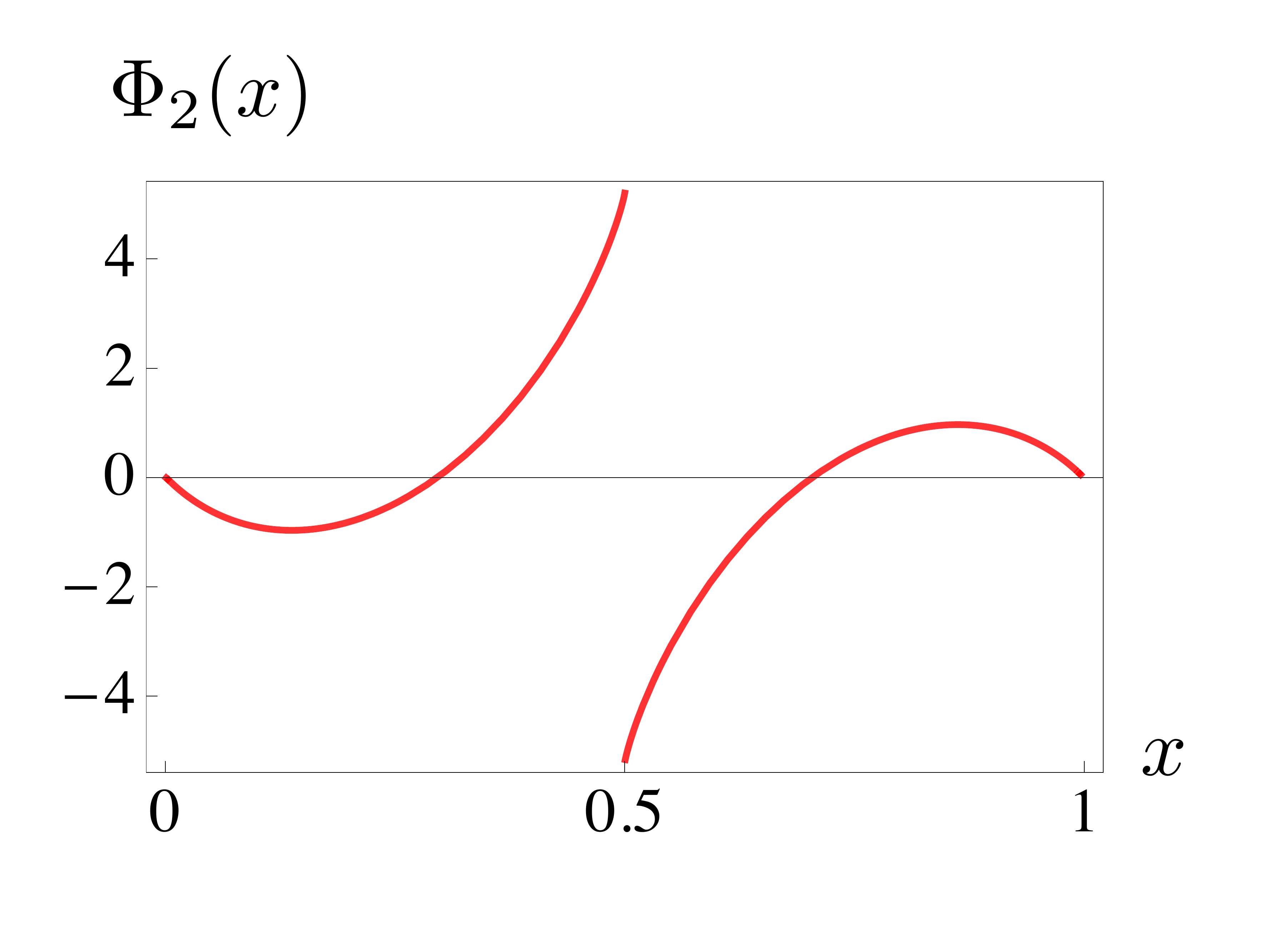}
\label{fig:subfig35}
}
\label{fig:subfigureExample}
\caption{ Expansion  components $\Phi_1(x)$  (a) and $\Phi_2(x)$  (b) in the case of the full kernel. }
\end{figure}

The distribution amplitude is  now  built using   
\begin{eqnarray}
\varphi(x,t)&=& e^{3/2t} (x\bar{x})^t|1-2x|^{2t}\left(\varphi_0(x)+
t \Phi_1(x)+\frac{t^2}{2!}\Phi_2(x)\right),
\end{eqnarray}
 For the first coefficient we have 
\begin{eqnarray}
 \Phi_1(x)=\theta ( 0<x\leqslant 1/2) \bigl [ -2x \ln 2 - 
 2\ln\bar{x} - (\bar{x} \ln\bar{x} + x \ln x) \bigl ]  - \{x \to \bar x \} \ ,  
\end{eqnarray}
and for the second, 
\begin{align}
 \Phi_2(x)=&\theta ( 0<x\leqslant 1/2 )
\Biggl \{ \frac{\pi ^2}{2}  (x-2)+2\ln^2\left(\frac{\bar{x}}{1-2x}\right)+(3 x-2) \ln^2 2\nonumber\\ 
&+\left(\frac{13}{2}-3 x\right) \ln^2 \bar{x}+\ln(1-2 x) (4(3-2x)\ln 2 +2(5-2 x) \ln x)\nonumber\\ 
&+\ln \bar{x} (1-2 (1-3x)\ln 2+(3 x-5) \ln x)+x \left(4  \ln x \ln2 +\ln \left(\frac{4 x}{\bar{x}}\right)\right)
\nonumber\\ 
&-4 \bar{x} \text{Li}_2 \left(\frac{1-2x}{2 \bar{x}}\right)-2 (x-4) \text{Li}_2\left(\frac{1-2 x}{\bar{x}}\right)-2 (5+x) \text{Li}_2(x)\nonumber\\  &+2(5-2 x) \text{Li}_2(2 x)-(1+2 x) \text{Li}_2\left(-\frac{x}{\bar{x}}\right)
 \nonumber\\ &+\text{Li}_2\left(\frac{x^2}{\bar{x}^2}\right)+4 \text{Li}_2\left(-\frac{x}{1-2 x}\right)\Biggl \}
 - \{x \to \bar x \} \ . 
\end{align}
As may be seen from Fig.10, the resulting curves are rather close to those
obtained when only singular part of  the kernel was taken into account.

\begin{figure}[h!]
\centering
\includegraphics[width=9cm]{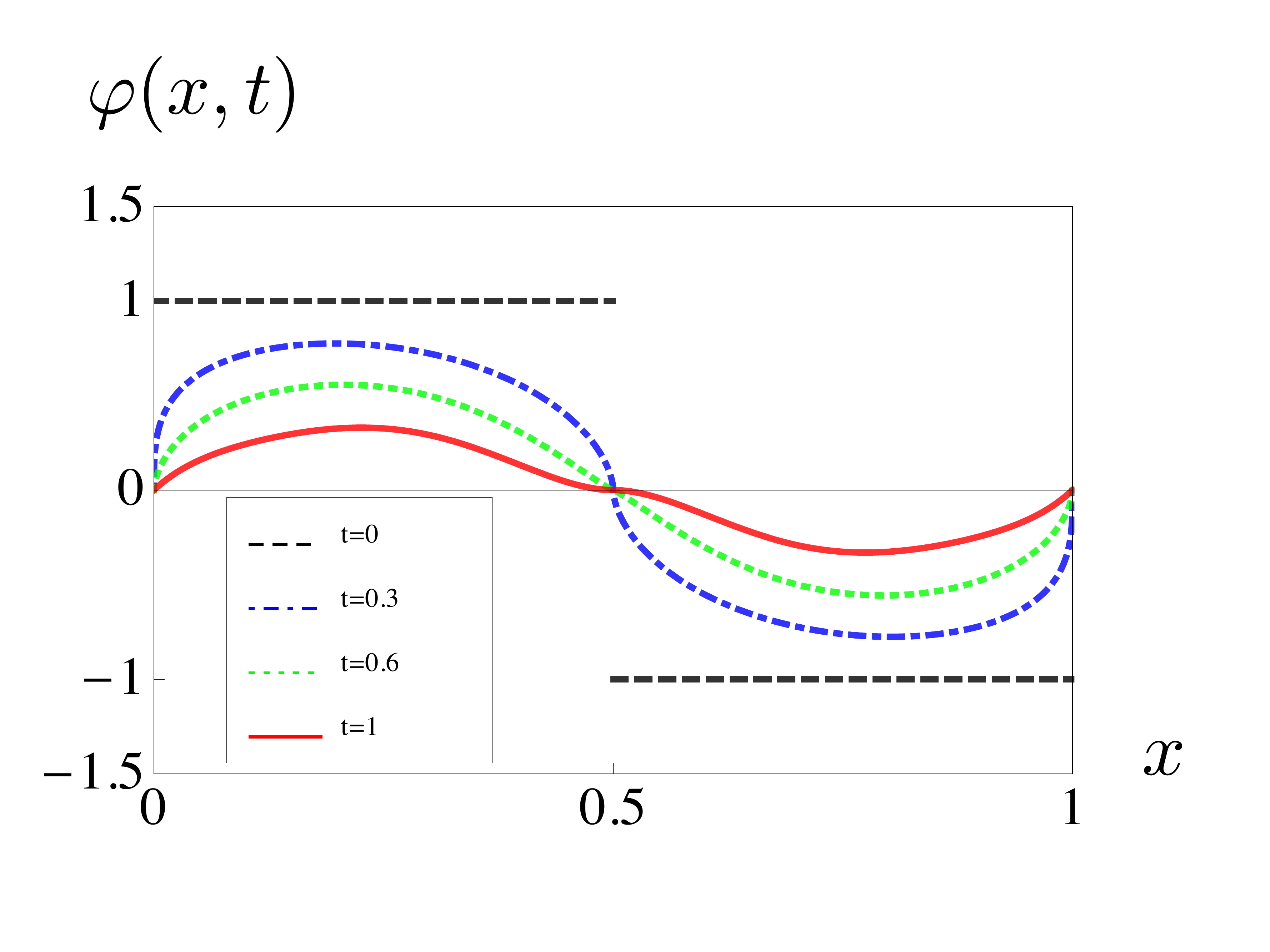}
\caption{ Evolution of  the antisymmetric DA in the  full kernel  case. 
The curves for $t=0,$  $ t=0.3, t=0.6, t=1$ are shown.}
\end{figure}


\section{Summary}

We described a  new analytical method of performing   ERBL evolution.
Unlike the standard method of expansion  in Gegenbauer polynomials, 
the method works for  functions that do not vanish at the end points.
The method was applied  for  two cases of the  initial DA:  for a purely flat DA,
constant in the whole $0\leq x \leq 1$ interval 
and for an antisymmetric DA which  is  constant in each of its two parts
$0 \leq x \leq 1/2$ and $1/2  \leq x \leq 1$.
In  case of  a purely flat DA, the leading term gives $(x \bar x)^t$  evolution
with the change of the evolution parameter $t$.
For the accompanying  factor, two further terms in the $t^N$  expansion 
were found.  In  case of  an antisymmetric flat DA,
there is an extra factor $|1-2x|^{2t}$ that takes care 
of the jump in the middle point $x=1/2$. 
The correction terms   were also calculated.
The results show good convergence for $t \lesssim 1/2$.
It  should  be noted  that for $t \gtrsim 1/2$,  the evolved DA 
is rather close to the asymptotic form, and one can use the standard 
method of the Gegenbauer expansion which is  well  convergent 
for such  functions.

Notice: Authored by Jefferson Science Associates, LLC under U.S. DOE Contract No.
 DE-AC05-06OR23177. The U.S. Government retains a non-exclusive, paid-up, 
 irrevocable, world-wide license to publish or reproduce this manuscript for U.S. Government purposes.



\begin{thebibliography}{1}

\bibitem{Aubert:2009mc}
B.~Aubert {\em et~al.}, {\em Phys. Rev.} {\bf D80},  052002 (2009).

\bibitem{Lepage:1980fj}
G.~P. Lepage and S.~J. Brodsky, {\em Phys. Rev.} {\bf D22},  2157 (1980).

\bibitem{Radyushkin:2009zg}
A.~V. Radyushkin, {\em Phys. Rev.} {\bf D80},  094009 (2009).

\bibitem{Polyakov:2009je}
M.~V. Polyakov, {\em JETP Lett.} {\bf 90}, 228 (2009).

\bibitem{Efremov:1979qk}
A.~V. Efremov and A.~V. Radyushkin, {\em Phys. Lett.} {\bf B94}, 245 (1980).

\bibitem{Brodsky:2011yv}
S.~J. Brodsky, F.-G. Cao and G.~F. de~Teramond, arXiv:1104.3364 [hep-ph] (2011).

\bibitem{Broniowski:2007si}
W.~Broniowski, E.~R. Arriola and K.~Golec-Biernat, {\em Phys. Rev.} {\bf D77},
   034023 (2008).

\bibitem{Polyakov:1999gs}
M.~V. Polyakov and C.~Weiss, {\em Phys. Rev.} {\bf D60},  114017 (1999).

\end{thebibliography}


\end{document}